\documentclass[lettersize,journal]{IEEEtran}
\usepackage{amsmath,amsfonts}
\usepackage{algorithmic}
\usepackage{array}
\usepackage[caption=false,font=normalsize,labelfont=sf,textfont=sf]{subfig}
\usepackage{textcomp}
\usepackage{stfloats}
\usepackage{url}
\usepackage{verbatim}
\usepackage{graphicx}
\usepackage{cite}
\usepackage{multirow}
\usepackage{float}
\usepackage{amsfonts,amssymb}
\usepackage{amsthm}
\usepackage{tabularx}
\usepackage{bbding}
\usepackage{color,xcolor}
\usepackage{booktabs}
\usepackage{hyperref}
\usepackage{verbatim}
\usepackage[linesnumbered,lined,boxed,commentsnumbered,ruled]{algorithm2e}
\makeatletter
\renewcommand{\maketag@@@}[1]{\hbox{\m@th\normalsize\normalfont#1}}
\makeatother
\usepackage{pifont}%
\newcommand{\cmark}{\ding{51}}%
\newcommand{\xmark}{\ding{55}}%
\begin{document}
\title{Causal Self-Supervised Pretrained Frontend with Predictive Patterns for Speech Separation}


\author{Wupeng Wang, Zexu Pan, Xinke Li, Shuai Wang,~\IEEEmembership{Member,~IEEE}, and Haizhou Li, ~\IEEEmembership{Fellow,~IEEE}}

\maketitle

\begin{abstract}
Speech separation (SS) seeks to disentangle a multi-talker speech mixture into single-talker speech streams. Although SS can be generally achieved using offline methods, such a processing paradigm is not suitable for real-time streaming applications. Causal separation models, which rely only on past and present information, offer a promising solution for real-time streaming. However, these models typically suffer from notable performance degradation due to the absence of future context. In this paper, we introduce a novel frontend that is designed to mitigate the mismatch between training and run-time inference by implicitly incorporating future information into causal models through predictive patterns. The pretrained frontend employs a transformer decoder network with a causal convolutional encoder as the backbone and is pretrained in a self-supervised manner with two innovative pretext tasks: autoregressive hybrid prediction and contextual knowledge distillation. These tasks enable the model to capture predictive patterns directly from mixtures in a self-supervised manner. The pretrained frontend subsequently serves as a feature extractor to generate high-quality predictive patterns. Comprehensive evaluations on synthetic and real-world datasets validated the effectiveness of the proposed pretrained frontend.
\end{abstract}


\section{Introduction}
\label{sec:introduction}
\IEEEPARstart{B}{lind} source separation seeks to separate speech signals by speakers in a speech mixture, which is often referred to as the `cocktail party' problem~\cite{bronkhorst2000cocktail}. It emulates human's selective auditory attention ability and benefits many downstream applications, such as automatic speech recognition (ASR) systems.~\cite{wang2020voicefilter,gabrys2022voice, pan2022hybrid}, voice conversion~\cite{zhou2020multi,zhou2021seen}, and speaker recognition~\cite{rao2019target,xu2021target,wang2024overview}.

Traditional speech separation techniques, including independent component analysis (ICA)~\cite{lee1998independent,araki2004underdetermined,choi2005blind}, non-negative matrix factorization (NMF)~\cite{wang2014discriminative,weninger2014discriminative,virtanen2007monaural}, and beamforming~\cite{ wang2018spatial,ikram2002beamforming, yin2018multi} primarily leverage spatial and spectral cues to distinguish between isolated sources. Recently, deep learning-based separation approaches~\cite{luo2019conv, luo2020dual, wang2020robust, wang2021neural, subakan2021attention, zhao2023mossformer, wang2023tf} with permutation invariant training (PIT)~\cite{yu2017permutation} have achieved notable improvements in quality and robustness through advanced neural network architectures. Despite these impressive results, deploying these models in real-world applications, such as telecommunications and hearing aids, remains a critical challenge: these applications require low latency and high reliability, yet most non-causal methods rely on utterance-level processing, which is impractical for achieving real-time separation. 

Causal speech separation models, designed for low latency signal processing, are well-suited for industrial applications that require real-time operations. A typical example is the causal convolutional neural network (CNN) introduced by Luo et al.~\cite{luo2019conv}, which utilizes causal convolutional layers to extract target voices and applies cumulative layer normalization to stabilize input features. Building on this, Li et al.~\cite{li2022skim, li2023predictive} incorporate a recurrent neural network (RNN) to alternately update both local and global features. The partial causal network offers another low-latency solution. For instance, Libera et al.~\cite{libera2024resource} propose using a time-averaged inner-chunk embedding as the summary representation for causal inter-chunk blocks to improve the separation quality, while Veluri et al.~\cite{veluri2023real} employ encoder-decoder architecture with a stack of dilated causal convolution to extract the specified target sound within limited latency. Although these models achieve low latency inference, they often exhibit significant quality degradation compared to their non-casual counterparts, especially on out-of-domain test sets in real scenarios. This performance gap stems primarily from the inherent limited access to future information. In this paper, we refer to this issue as the `\emph{causal separation}' problem. 

Human auditory systems demonstrate remarkable capabilities in real-time speech separation. A general observation is when we enter a complex auditory environment, it can be challenging to distinguish a specific voice. However, as we acclimate to the surroundings, identifying the target voice becomes easier, which is known as the \textit{steady-state response} in the auditory cortex, inferior frontal gyrus, and hippocampus~\cite{szabo2016computational}. The theory of computational auditory scene analysis (CASA) suggests that this capacity arises from the ability of the auditory system to capture predictive patterns~\cite{barascud2016brain,kumar2016brain}, such as tones and phonemes. In~\cite{bregman1994auditory}, Bregman demonstrates the effect of tones through a perceptual continuation experiment of alternating tone sequences. Warren et al.~\cite{warren1970perceptual} illustrate that listeners perceive a spoken sentence as continuous despite interruptions by loud noise bursts, which is referred to as the \textit{phonemic restoration} phenomenon. Humans develop such listening skills during their childhood unconsciously through listening to large-scale speech interactions. We attempt to emulate these human capabilities via a pretrained frontend to boost the separation quality of causal separation models.

Pretrained frontends are learnable feature extractors that have emerged as a prominent paradigm in recent years. By using self-supervised learning (SSL) training methods, these frameworks demonstrate superior capability in learning high-level abstractions from unlabeled real-world scenarios speech, which is similar to the unconscious listening process of humans. As far as causality is concerned, the pretrained frontends can be divided into two categories: non-causal frontends~\cite{chen2022unispeech, chen2022wavlm, fazel2023cocktail, lin2024selective,liu2020mockingjay,liu2021tera,chi2021audio, wang2022wav2vec,hsu2021hubert,baevski2022data2vec,wang2023data2vec,zhu2023robust, hu2023wav2code, wang2024speech}, and causal frontends~\cite{chung2020generative, chung2020vector, oord2018representation, schneider2019wav2vec, baevski2019vq}. Non-causal frontends leverage the full utterance to extract high-level contextual representations, while causal frontends process only past and current information. Both causal and non-causal frontends have demonstrated considerable improvements across various single-speaker speech tasks~\cite{chen2023speech}, such as automatic speech recognition, and speaker identification. However, there are limited explorations of causal frontends specifically designed for real-scenario causal speech separation, where the input is multi-speaker mixtures without target reference speech.

In this paper, we employ a transformer decoder network\cite{Vaswani2017attention} with a causal convolutional encoder as the causal self-supervised pretrained (CSP) frontend to overcome the `\emph{causal separation}' problem. The primary contributions of this paper are summarized as follows,

\begin{itemize}
    \item We propose the first causal self-supervised pretrained frontend specifically designed for the multi-speaker real-world scenarios to improve the separation quality of causal speech separation models;
    \item We introduce the concept of leveraging mixture waveforms for self-supervised frontend training and propose the autoregressive hybrid prediction (AHP) pretext task to capture predictive patterns; 
    \item We develop the contextual knowledge distillation (CKD) pretext task to facilitate the frontend to encode contextual information into predictive patterns;
    \item We conduct comprehensive experiments to demonstrate the superior results of using predictive patterns across diverse real and synthetic datasets, with various causal separation models and evaluation metrics.    
\end{itemize}

The rest of the paper is organized as follows. Section~\ref{sec:related_work} discusses the latest developments in pretrained frontends. Section~\ref{sec:model} formulates the pretraining method and introduces the architecture of our CSP frontend. In Section~\ref{sec:exp_setup}, we describe the model configuration and experimental setup. Section~\ref{sec:exp_result} reports benchmark comparisons of leading techniques and provides a detailed analysis of performance metrics. Finally, Section~\ref{sec:con_ssl} concludes the paper.

\section{Related Work on pretrained Frontend}
\label{sec:related_work}

The pursuit of effective pretrained frontends is essential to address the growing demands for high-quality audio processing. According to their temporal dependencies, these frontends can be categorized into two types: non-causal frontends that utilize both past and future context, and causal frontends that operate solely on historical information. We first review representative non-causal frontends in Sections~\ref{sec:MAM} and~\ref{sec:MPP}, followed by a detailed discussion of causal frontends in Sections.~\ref{sec:APC} and~\ref{sec:CPC}.

\subsection{Masked Acoustic Modeling Frontend}
\label{sec:MAM}
The masked acoustic modeling (MAM) frontend is one of the most popular non-causal frontends in recent years. It forces the frontend to generate a probability distribution of the entire sequence through a fill-in-the-blank task with one or multiple missing elements, which is similar to the well-performing masked language modeling (MLM)~\cite{devlin2018bert} pretraining strategy. During the pretraining stage, the frontend reconstructs the randomly masked linear-scale spectrogram using Euclidean distance loss or recognizes the latent features of the input waveform through noise-contrastive
estimation (NCE)~\cite{mnih2012fast,gutmann2010noise,jozefowicz2016exploring}. 

In~\cite{liu2020mockingjay}, the Mockingjay frontend maps the input into log Mel-spectrogram and randomly masks a specific portion of the frames for a stacked transformer encoder network to predict masked clips. Tera~\cite{liu2021tera} incorporates both time and frequency masking with various mask ratios to promote the robustness of the frontend. Instead of predicting the spectrogram, Wav2vec2.0~\cite{baevski2020wav2vec} learns high-level slow features by recognizing the positive quantized code from negative candidates through mutual information maximization~\cite{oord2018representation}. Recently, Data2Vec~\cite{wang2023data2vec} introduces a universal teacher-student learning mechanism with an exponential moving average update method to estimate predictive patterns. Although these MAM frontends perform well on various speech tasks, the non-causal property of these frontends limits their use to real-time streaming applications. 

\subsection{Masked Pseudo-label Prediction  Frontend}
\label{sec:MPP}
Instead of predicting masked clips as the MAM frontend, the masked pseudo-label prediction (MPP) frontend provides an alternative approach for learning high-level phoneme-related information on the non-causal frontend. The MPP frontend, introduced in Hubert~\cite{hsu2021hubert}, collects pseudo-labels from the raw waveform through unsupervised clustering algorithms on the mel frequency cepstral coefficient (MFCC) feature. A transformer network is then employed to predict these pseudo-labels with masked inputs. In this way, the frontend learns to discover the hidden units that are related to the phoneme information in a self-supervised manner. The authors also demonstrate that this learning procedure can be repeated multiple times. 

Recently, numerous variants of Hubert have emerged. In~\cite{chen2022unispeech}, Chen et al. propose a primary speaker denoising (PSD) pretext task to learn robust hidden units. The authors collect frame-level cluster labels from target reference speech and train the frontend to predict these pseudo-labels from synthetic mixtures generated by utterance mixing augmentation. WavLM~\cite{chen2022wavlm} extends this framework with relative positional encoding and verifies the effectiveness of the frontend in various downstream tasks. To further explore the utilization of the frontend in speech separation tasks, Wang et al.~\cite{wang2023adapter,fazel2023cocktail, ng2023hubert, lin2024selective} design various weakly-supervised pseudo-label prediction tasks to capture the hidden unit information of all target speakers in the mixture waveform. Despite convincing improvements in numerous downstream tasks, the lack of parallel mixtures and their corresponding target reference speech makes it difficult to implement in real-world scenarios. 

\subsection{Autoregressive Predictive Coding Frontend}
\label{sec:APC}
 Unlike non-causal frontends, the causal frontend captures predictive patterns without access to future information. The autoregressive predictive coding (APC) frontend is one of the causal pretrained frontends that learns representations from the sequential spectrogram. It assumes that the probability distribution of the spectrogram at a specific time step is conditionally dependent on the preceding information in the raw input sequence. From this perspective, the frontend leverages the autoregressive method to predict the future spectrogram of a sequence based on its past by minimizing the feature distance, and thus capturing the long-range dependency of the raw audio waveform.
 
 The effectiveness of the APC frontend has been widely explored in single-speaker phoneme classification and ASR tasks. In~\cite{chung2020generative}, Chung et al. first investigated using stacked long short-term memory (LSTM) recurrent neural networks to learn speech representations. By predicting the Mel spectrogram of future frames, the APC frontend demonstrated significant improvement in phoneme classification tasks. Its successor, the VQ-APC~\cite{chung2020vector}, further extended this method through vector quantization and multi-task training to discover discrete speech units for phoneme recognition and ASR tasks. However, as reported in~\cite{tsai2022superb}, the APC frontend may not be suitable for speech separation, as it disregards phase information, which is important for waveform reconstruction.

\subsection{Contrastive Predictive Coding Frontend}
\label{sec:CPC}
Rather than directly reconstructing the future spectrogram, the contrastive predictive coding (CPC) frontend aims to capture high-level context that spans multiple steps in a discriminative manner using noise-contrastive estimation (NCE)~\cite{mnih2012fast,gutmann2010noise,jozefowicz2016exploring}. Typically, the CPC frontend encodes various levels of abstraction into compact contextual features that maximally preserve the mutual information between observations and their latent representations. 

The first exploration of the CPC frontend is designed by Oord et al.~\cite{oord2018representation}, where a convolutional neural network (CNN) is used to map the audio waveform into latent space, followed by a stacked recurrent neural network (RNN) to capture global abstraction via mutual information maximization. In~\cite{schneider2019wav2vec}, Schneider et al. extend this idea by replacing the RNN network with a convolution structure and demonstrating its effectiveness in ASR downstream tasks. In VQ-Wav2vec~\cite{baevski2019vq}, Baevski et al. further utilize a trainable discrete quantization module to integrate the frontend with acoustic models. In contrast to other frontends, the CPC frontend is effective for extracting predictive patterns, as it captures multi-scale abstractions from input waveforms without discarding essential signal information~\cite{pasad2021layer}.

\section{Causal Frontend Pretrained on Speech Mixtures}
\label{sec:model}

As discussed in Section.~\ref{sec:introduction}, the key to overcoming the \emph{`causal separation'} problem hinges on leveraging predictive patterns that encapsulate future information. Recently, extensive efforts have been made to extract these patterns from large-scale single-speaker waveforms, primarily through pretraining methods~\cite{chung2020generative, chung2020vector, oord2018representation, schneider2019wav2vec, baevski2019vq}. However, these methods are less effective for multi-speaker mixtures~\cite{huang2022investigating}, especially in real-world scenarios where individual target reference speech is unavailable.

In this section, we introduce a novel causal self-supervised pretrained (CSP) frontend with two pretext tasks for capturing predictive patterns from multi-speaker mixtures. The first pretext task, autoregressive hybrid prediction (AHP), enables the frontend to extract predictive patterns through an autoregressive framework as detailed in Section.~\ref{sec:AHP}. The second pretext task, contextual knowledge distillation (CKD), facilitates the frontend to encode contextual information into predictive patterns by transferring knowledge from a non-causal teacher model to a causal student model, as demonstrated in Section.~\ref{sec:CKD}. The multi-task loss and the overall architecture of the proposed frontend are illustrated in Section.~\ref{sec:MT} and Section.~\ref{sec:structure}, respectively.

\begin{figure*}[t]
	\centering 
	\includegraphics[width=1.0\linewidth]{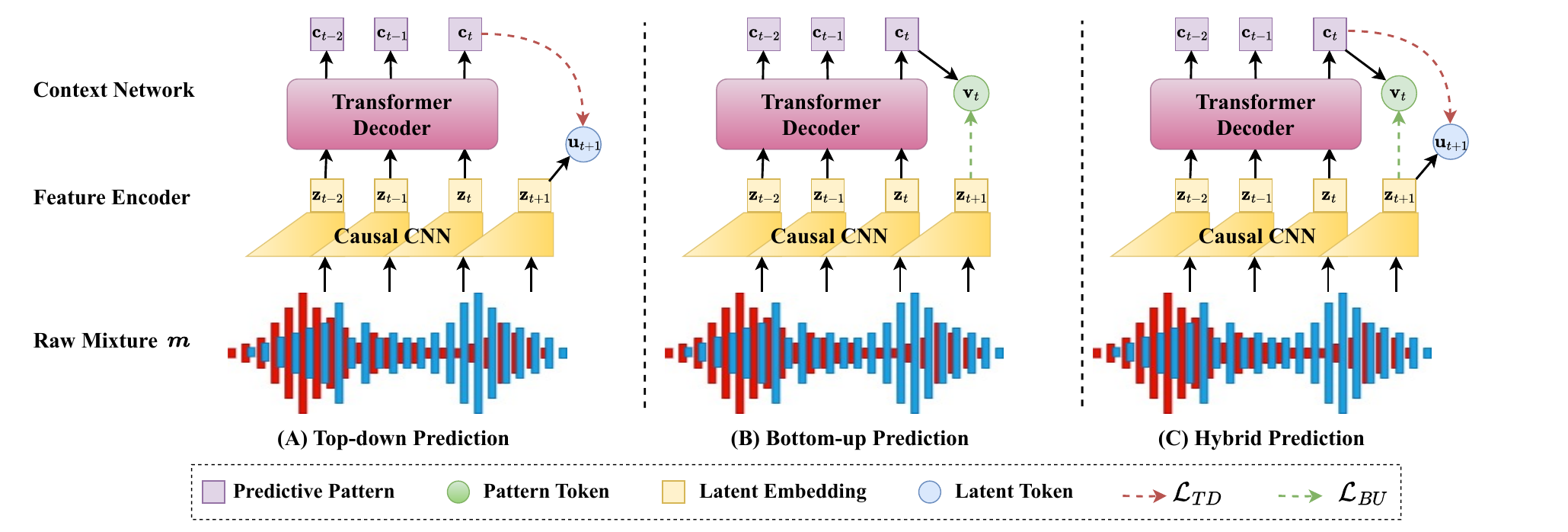}
	\caption{The details of AHP pretext task. $\mathbf{z}_{t+1}$ is the future mixture latent embedding. $\mathbf{c}_{t}$ is the current predictive pattern. $\mathbf{u}_{t+1}$ is future latent token, while $\mathbf{v}_{t}$ is the current pattern token. The $\mathcal{L}_{\text{TD}}$ and $\mathcal{L}_{\text{BU}}$ are loss functions defined at Eq.~\eqref{equ.ahp_td} and Eq.~\eqref{equ.ahp_bu}, respectively.} 
	\label{Fig.AHP} 
\end{figure*}

\subsection{Autoregressive Hybrid Prediction}
\label{sec:AHP}

Mutual information (MI) is a fundamental metric for quantifying the nonlinear statistical dependence between random variables and is widely employed to capture predictive patterns in speech waveforms. In~\cite{oord2018representation}, Oord et al. first apply the mutual information to single-speaker speech pretraining, by maximizing the mutual information between current predictive pattern and future signals. We extend this approach as the autoregressive hybrid prediction (AHP) pretext task to discover predictive patterns in a cocktail party without corresponding reference speech.

Let $\mathcal{D}_m$ denote an unlabeled multi-speaker mixture waveform dataset. For an arbitrary mixture sample $\boldsymbol{m} \in \mathbb{R}^{\tau \times 1}$ in $\mathcal{D}_m$, we assume that all its isolated target reference speech are additively combined. Our goal is to generate predictive pattern sequence $\mathbf{C} = \left[\mathbf{c}_0, \mathbf{c}_{1}, \ldots, \mathbf{c}_{T-1}, \mathbf{c}_T\right]$\footnote{In this paper, a variable with $\tau$ as the index represents a sequence of samples
in the time domain, while a variable with $T$ as the index represents a sequence
of latent embeddings.} using our proposed CSP frontend shown in Fig.~\ref{Fig.AHP}. Following the suggestion in~\cite{oord2018representation}, the current predictive pattern $\mathbf{c}_t$ at time $t$ should maximize the mutual information $I(\mathbf{c}_t;\mathbf{s}_{t+1}^i)$ between $\mathbf{c}_t$ and the future reference speech latent embedding $\mathbf{s}_{t+1}^i$ from each speaker $i$ at time $t+1$. However, since the target reference speech is unavailable for real mixtures, directly discover the predictive pattern $\mathbf{c}_t$ by maximizing $I(\mathbf{c}_t;\mathbf{s}_{t+1}^i)$ is not feasible. Assume that $\mathbf{Z} = \left[\mathbf{z}_0, \mathbf{z}_{1}, \ldots, \mathbf{z}_{T-1}, \mathbf{z}_T\right]$ denotes the latent embedding sequence of the mixture $\boldsymbol{m}$ extracted by the causal encoder of the frontend. We exploit the lower bound of $I(\mathbf{c}_t;\mathbf{s}_{t+1}^i)$ as follows:

\begin{equation}
    \resizebox{1.0\columnwidth}{!}{$\displaystyle{
    \begin{split}
        I\left(\mathbf{c}_t; \mathbf{s}_{t+1}^i\right) & = H\left(\mathbf{s}_{t+1}^i\right) - H\left(\mathbf{s}_{t+1}^i|\mathbf{c}_t\right)\\
        & \geq \underbrace{\left[H\left(\mathbf{s}_{t+1}^i\right) - H\left(\mathbf{z}_{t+1}\right)\right] + \left(H\left(\mathbf{z}_{t+1}\right) - H\left(\mathbf{z}_{t+1}|\mathbf{c}_t\right)\right)}_{\text{Lower bound}}\\
        & = \underbrace{I\left(\mathbf{c}_t;\mathbf{z}_{t+1}\right)}_{\text{Mixture MI}} + \underbrace{\left[H\left(\mathbf{s}_{t+1}^i\right)-H\left(\mathbf{z}_{t+1}\right)\right]}_{\text{Constant}}
    \end{split}
    }$}
    \label{equ.pred_mi}
\end{equation}

Here $\mathbf{z}_{t+1}$ is the future mixture latent embedding at time $t+1$. The left-hand side of Eq.~\eqref{equ.pred_mi} represents the mutual information we seek to maximize, while the right-hand side provides its lower bound. This lower bound arises from the principle that the entropy of the sum of two independent variables is at least as large as the maximum entropy of its individual components. The first component $I\left(\mathbf{c}_t;\mathbf{z}_{t+1}\right)$ quantifies the mixture mutual information between the current predictive pattern $\mathbf{c}_t$ at time $t$ and the future latent embedding $\mathbf{z}_{t+1}$ of mixture speech at time $t+1$. The second component $\left[H\left(\mathbf{s}_{t+1}^i\right)-H\left(\mathbf{z}_{t+1}\right)\right]$ is a constant value denoting the entropy difference between the mixture latent embedding $\mathbf{z}_{t+1}$ and the target reference speech latent embedding $\mathbf{s}_{t+1}^i$ at time $t+1$. Eq.~\eqref{equ.pred_mi} illustrates that maximizing $I\left(\mathbf{c}_t; \mathbf{z}_{t+1}\right)$ effectively maximizes the lower bound of $I\left(\mathbf{c}_t; \mathbf{s}_{t+1}^i\right)$, thereby encoding the future information of reference speech into predictive patterns. To exploit this principle, we employ both top-down and bottom-up prediction, as well as their hybrid version, to extract predictive patterns from unlabeled mixtures through self-supervised training.

\subsubsection{Top-down prediction}
\label{sec:TD}
We first follow the NCE method~\cite{oord2018representation} to maximize the $I\left(\mathbf{c}_t;\mathbf{z}_{t+1}\right)$. As demonstrated in the Fig.~\ref{Fig.AHP} (A), We randomly select a normalized mixture waveform, denoted as $\boldsymbol{m}$, from the unlabeled multi-speaker mixture dataset $\mathcal{D}_m$. Then, we extract the current latent embedding sequence and the future latent embedding $\mathbf{z}_{t+1}$ through the feature encoder. $\mathbf{z}_{t+1}$ is then quantized with the embedding quantization module to discretize as the future latent token $\mathbf{u}_{t+1}$. Next, a random mask is applied to the current latent embedding sequence with a specific ratio $\delta$ to obtain the masked latent embedding sequence. Finally, the masked embedding sequence is fed into the context network to estimate the predictive pattern $\mathbf{c}_t$. We pretrain the frontend by maximizing the $I(\mathbf{c}_t;\mathbf{u}_{t+1})$ with $\mathcal{L}_{\text{TD}}$ loss function to optimize the frontend:

\begin{equation}
    \label{equ.ahp_td}
    \mathcal{L}_{\text{TD}} = \mathcal{L}_{u} + \mathcal{L}_{\text{TD-NCE}}\left(\mathbf{c}_t,\mathbf{u}_{t+1}\right),
\end{equation}
where $\mathcal{L}_{u}$ is the diversity loss of the embedding quantization module, which contains $G_{u}$ quantization codebooks with $R_{u}$ entries, to encourage the usage of all quantized codes~\cite{baevski2020wav2vec}: 

\begin{equation}
    \label{equ.td_q}
    \mathcal{L}_{u} = -\log \frac{1}{G_{u}R_{u}}\sum_{g=1}^{G_{u}}\sum_{r=1}^{R_{u}}p_{g,r}\log p_{g,r}
\end{equation}

Here $p_{g,r}$ is the selected probability of the $r^{th}$ quantized code in the $g^{th}$ codebook. The $\mathcal{L}_{\text{TD-NCE}}$ with temperature $\omega$ and similarity function $\psi$ serves as the contrastive loss to distinguish the future latent token $\mathbf{u}_{t+1}$ from a set of latent candidates $\mathcal{Q}_U$, which includes $\mathbf{u}_{t+1}$ and $N$ negative distractors $\tilde{\mathbf{u}}$:

\begin{equation}        
    \label{equ.td_nce}
    \mathcal{L}_{\text{TD-NCE}}\left(\mathbf{c}_t,\mathbf{u}_{t+1}\right) = -\log \frac{\exp\left(\psi\left(\mathbf{c}_t,\mathbf{u}_{t+1}\right) / \omega\right)}{\sum_{\tilde{\mathbf{u}} \in \mathcal{Q}_U} \exp\left(\psi\left(\mathbf{c}_t,\mathbf{\tilde{u}}\right) / \omega\right)}
\end{equation}

\subsubsection{Bottom-up prediction}
\label{sec:BU}
While top-down prediction is well-suited for capturing predictive patterns, it inherently introduces the `\textit{output independence hypothesis}': the current latent embedding sequence is independent of the future latent embedding $\mathbf{z}_{t+1}$ when the predictive pattern $\mathbf{c}_t$ is given, which violates speech coherence property. To address this limitation, we propose a bottom-up prediction as the complementary to ensure the continuity of the waveform.

Our bottom-up prediction method is shown in Fig.~\ref{Fig.AHP} (B), which is inspired by the maximum-entropy Markov model (MEMM). We follow the top-down prediction procedure to extract the future latent embedding $\mathbf{z}_{t+1}$, and the current predictive pattern $\mathbf{c}_{t}$, respectively. Due to the symmetry of $I\left(\mathbf{c}_t;\mathbf{z}_{t+1}\right)$, we quantize $\mathbf{c}_{t}$ with the pattern quantization module as the current pattern token $\mathbf{v}_{t}$ and utilize $\mathbf{z}_{t+1}$ to identify it. We utilize a bottom-up loss $\mathcal{L}_{\text{BU}}$ to pretrain the fontend through maximizing $I(\mathbf{v}_{t};\mathbf{z}_{t+1})$. The loss function is defined as follows:

\begin{equation}
    \label{equ.ahp_bu}
    \mathcal{L}_{\text{BU}} = \mathcal{L}_{v} + \mathcal{L}_{\text{BU-NCE}}\left(\mathbf{v}_{t},\mathbf{z}_{t+1}\right)
\end{equation}

Here the $\mathcal{L}_{v}$ is the diversity loss of the pattern quantization module, which contains $G_{v}$ quantization codebooks with $R_{v}$ entries: 

\begin{equation}
    \label{equ.bu_q}
    \mathcal{L}_{v} = -\log \frac{1}{G_{v}R_{v}}\sum_{g=1}^{G_{v}}\sum_{r=1}^{R_{v}}p_{g,r}\log p_{g,r}
\end{equation}

Similarly, the $p_{g,r}$ here is the selected probability of the $r^{th}$ quantized code in the $g^{th}$ codebook. The $\mathcal{L}_{
\text{BU-NCE}}$ is the contrastive loss for latent embedding $\mathbf{z}_t$ to recognize the pattern token $\mathbf{v}_{t}$ from a set of predictive candidates $\mathcal{Q}_V$ with positive sample $\mathbf{v}_{t}$ and N negative distractors $\tilde{\mathbf{v}}$:

\begin{equation}        
    \label{equ.bu_nce}
    \mathcal{L}_{\text{BU-NCE}}\left(\mathbf{v}_{t}, \mathbf{z}_{t+1}\right) = -\log \frac{\exp\left(\psi\left(\mathbf{v}_{t}, \mathbf{z}_{t+1}\right) / \omega\right)}{\sum_{\mathbf{\tilde{v}} \in \mathcal{Q}_V} \exp\left(\psi\left(\tilde{\mathbf{v}}, \mathbf{z}_{t+1}\right) / \omega\right)}
\end{equation}


\subsubsection{Hybrid prediction}
Although the bottom-up prediction task preserves speech coherence, the convolutional encoder lacks sufficient capacity to effectively encode the information of speech mixtures. We introduce the autoregressive hybrid prediction (AHP) pretext task to combine both bottom-up and top-down predictions together. The loss function for the AHP pretext task is defined as follows:

\begin{equation}
    \label{equ.ahp_hybrid}
    \mathcal{L}_{\text{AHP}} = \alpha \times \mathcal{L}_{\text{TD}} + \beta \times \mathcal{L}_{\text{BU}}
\end{equation}

Here, the hyper-parameters $\alpha$ and $\beta$ are the balance weights between top-down and bottom-up predictions.

\subsection{Contextual Knowledge Distillation}
\label{sec:CKD}

In the AHP pretext task, we maximize the $I\left(\mathbf{c}_t;\mathbf{z}_{t+1}\right)$ to encode the future information into predictive patterns. However, as noted by~\cite{szabo2016computational}, effective predictive patterns should also capture high-level contextual abstractions. To adapt our causal frontend for this capability, we introduce contextual knowledge distillation (CKD) pretext task to pretrain the frontend in a self-supervised manner. Assume that the $\mathbf{\overline{C}} = \left[\mathbf{\overline{c}}_0, \mathbf{\overline{c}}_{1}, \ldots, \mathbf{\overline{c}}_{T-1}, \mathbf{\overline{c}}_T\right]$ denote the contextual representation sequence of mixture $\boldsymbol{m}$, we extend optimization target $I\left(\mathbf{c}_t;\mathbf{s}_{t+1}^i\right)$ by incorporating the contextual abstraction $\mathbf{\overline{c}}_{t}$ through the chain rule of mutual information as follows:

\begin{equation}
    \begin{split}
        I\left(\mathbf{c}_t; \mathbf{s}_{t+1}^i\right) 
        & = I\left(\mathbf{c}_t; \mathbf{\overline{c}}_{t}, \mathbf{s}_{t+1}^i\right) - I\left(\mathbf{c}_t;\mathbf{\overline{c}}_{t}\mid\mathbf{s}_{t+1}^i\right)\\
        & = I\left(\mathbf{c}_t; \mathbf{\overline{c}}_{t}\right) + I\left(\mathbf{c}_t; \mathbf{s}_{t+1}^i\mid\mathbf{\overline{c}}_{t}\right)\\ & \quad - I\left(\mathbf{c}_t;\mathbf{\overline{c}}_{t}\mid\mathbf{s}_{t+1}^i\right)\\
        & \geq I\left(\mathbf{c}_t; \mathbf{\overline{c}}_{t}\right) - I\left(\mathbf{c}_t;\mathbf{\overline{c}}_{t}\mid\mathbf{s}_{t+1}^i\right)\\
        & = I\left(\mathbf{c}_t; \mathbf{\overline{c}}_{t}\right) - H\left(\mathbf{\overline{c}}_{t}\mid\mathbf{s}_{t+1}^i\right) + H\left(\mathbf{\overline{c}}_{t}\mid\mathbf{c}_t;\mathbf{s}_{t+1}^i\right)\\
        & \geq \underbrace{I\left(\mathbf{c}_t; \mathbf{\overline{c}}_{t}\right)}_{\text{Contextual MI}} - \underbrace{H\left(\mathbf{\overline{c}}_{t}\mid\mathbf{s}_{t+1}^i\right)}_{\text{Constant}}
    \end{split}
    \label{equ.chain}
\end{equation}

\begin{figure}[t]
	\centering 
	\includegraphics[width=1.0\linewidth]{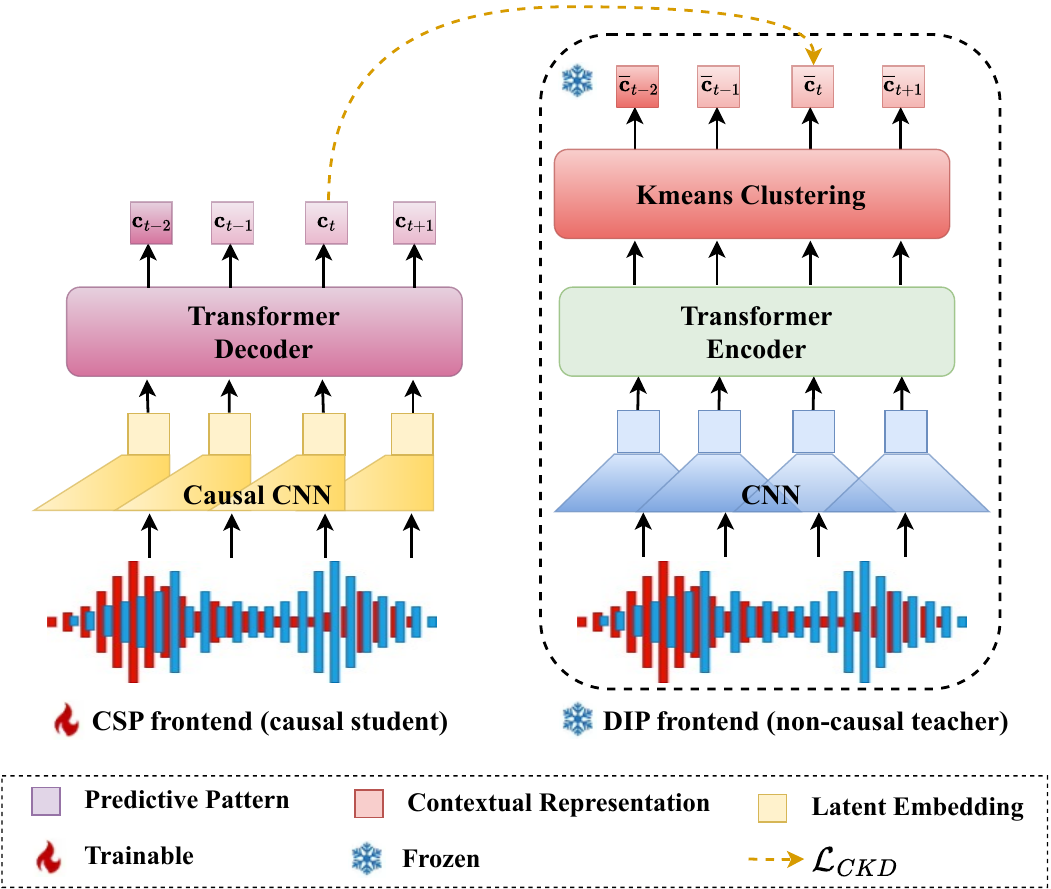}
	\caption{The details of CKD pretext task. The CSP frontend is the trainable causal student model, while the DIP frontend is the frozen non-causal teacher model. The $\mathbf{c}_{t}$ is the current predictive pattern. The $\overline{\mathbf{c}}_{t}$ is the current contextual representation. The $\mathcal{L}_{\text{CKD}}$ is the distillation loss in Eq.~\eqref{equ.ckd}.} 
	\label{Fig.CKD} 
\end{figure}

The left-hand side of Eq.~\eqref{equ.chain} represents the optimization target, while the right-hand side provides its another lower bound. The first component $I\left(\mathbf{c}_t; \mathbf{\overline{c}}_{t}\right)$ quantifies the contextual mutual information between the predictive pattern $\mathbf{c}_t$ and the contextual representation $\mathbf{\overline{c}}_t$ at time $t$. The second component $H\left(\mathbf{\overline{c}}_{t}\mid\mathbf{s}_{t+1}^i\right)$ is a constant value that indicates the relationship between context $\mathbf{\overline{c}}_t$ and its future observation $\mathbf{s}_{t+1}^i$, which is independent of the predictive pattern $\mathbf{c}_t$.  Eq.~\eqref{equ.chain} illustrates that maximizing $I\left(\mathbf{c}_t; \mathbf{\overline{c}}_{t}\right)$ effectively maximizes the lower bound of $I\left(\mathbf{c}_t; \mathbf{s}_{t+1}^i\right)$, thereby encoding the contextual information into predictive patterns. In~\cite{wang2024speech}, Wang et al. demonstrate that the $\mathbf{\overline{c}}_{t}$ can be extracted through the non-causal domain-invariant pretrained (DIP) frontend. Therefore, we treat the DIP frontend as a frozen teacher model and propose a contextual knowledge distillation (CKD) pretext task to facilitate our CSP frontend to capture contextual information. The pretraining procedure of the CKD is illustrated in Fig.~\ref{Fig.CKD}.  

First, we collect the output from the non-causal DIP frontend and utilize the K-means clustering method to generate the $K$ cluster centers. The cluster center embedding $\mathbf{\overline{c}}_{t}$ at time $t$ is then used as the contextual representation for the knowledge distillation process. Next, we follow the AHP pretext task to extract the current predictive pattern $\mathbf{c}_{t}$ at the same time $t$. Finally, we facilitate knowledge transfer by minimizing the $I\left(\mathbf{c}_t; \mathbf{\overline{c}}_{t}\right)$ between the contextual representations $\mathbf{\overline{c}}_{t}$ from the non-causal teacher model and the predictive pattern $\mathbf{c}_{t}$ from the causal student model through a knowledge distillation loss $\mathcal{L}_{\text{CKD}}$ defined as follows: 

\begin{equation}        
    \label{equ.ckd}
    \mathcal{L}_{\text{CKD}}\left(\mathbf{c}_{t}, \mathbf{\overline{c}}_{t}\right) = -\log \frac{\exp\left(\psi\left(\mathbf{c}_{t}, \mathbf{\overline{c}}_{t}\right) / \omega\right)}{\sum_{\tilde{\mathbf{c}} \in \mathcal{Q}_{\overline{c}}} \exp\left(\psi\left(\mathbf{c}, \tilde{\mathbf{c}}\right) / \omega\right)},
\end{equation}
where $\mathcal{Q}_{\overline{c}}$ is a set of candidates that includes positive sample $\mathbf{\overline{c}}_{t}$ and uniformly sampled negative distractors $\tilde{\mathbf{c}}$ from the non-causal teacher model. 

\subsection{Multi-task Loss}
\label{sec:MT}
The AHP pretext task facilitates the frontend to extract predictive patterns from mixtures, while the CKD pretext task enables the patterns to encode contextual information. To ensure the frontend simultaneously captures both aspects, we integrate these tasks within a unified multi-task loss framework as follows:

\begin{equation}        
    \label{equ.CSP}
    \mathcal{L}_{\text{CSP}} =\mathcal{L}_{\text{AHP}} + \gamma \times \mathcal{L}_{\text{CKD}},
\end{equation}
where $\gamma$ is a hyper-parameter to balance the contribution of two pretext tasks.

\begin{figure}[t]
	\centering 	\includegraphics[width=1.0\linewidth]{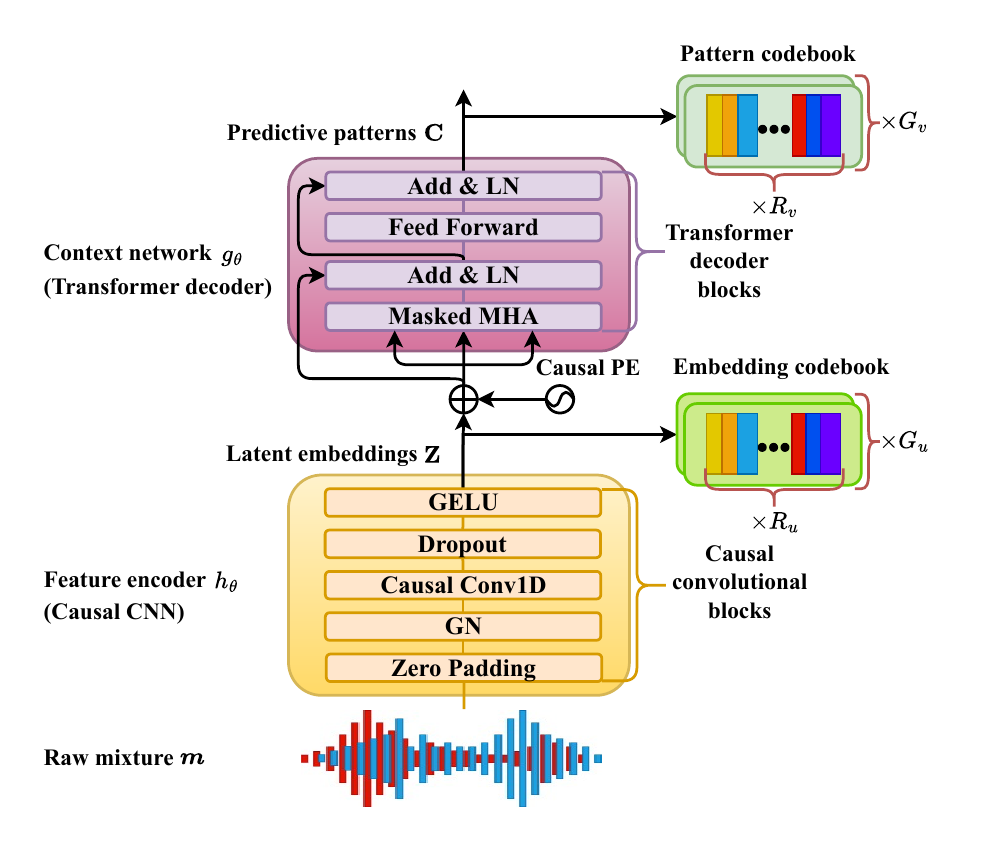}
\caption{The architecture of CSP frontend. The causal Conv1D is the causal 1-dimensional convolutional layer. The GN and LN are the frame-wise group normalization and layer normalization. The GELU is the Gaussian error linear unit activation function. The causal PE is the causal learnable positional encoding. The MHA is the multi-head attention layer.} 
	\label{Fig.structure} 
\end{figure}

\subsection{Network Structure}
\label{sec:structure}
The overall structure of the frontend is illustrated in Fig.~\ref{Fig.structure}.

\subsubsection{Feature Encoder}
The feature encoder is a pyramidal causal CNN structure to capture latent embeddings sequence $\mathbf{Z}$ from raw mixture waveform $\boldsymbol{m}$ with spectral characteristics. It comprises stacked causal convolutional blocks, each of which is followed by a dropout layer and GELU activation functions~\cite{hendrycks2016gaussian}. To preserve causality, zero-padding is applied to the input of each block. Frame-wise group normalization~\cite{wu2018group} is applied to the initial block to stabilize feature scaling.  

\subsubsection{Context network}
The context network is designed to generate predictive pattern sequence $\mathbf{C}$ in an autoregressive manner with the transformer decoder network. It contains stacked transformer decoder blocks with a learnable causal positional embedding to capture temporal relationships among input features. To improve stability during pretraining, the network applies a GELU activation function and layer normalization for each transformer block.

\subsubsection{Quantization module}
The embedding quantization module applies $G_u$ codebooks to discretize the latent embedding into $R_u$ latent token using product quantization. Likewise, the pattern quantization module utilizes $G_v$ codebooks to convert the predictive pattern into $R_v$ pattern token. A Gumbel-softmax function is incorporated to enable fully differentiable backpropagation.

\section{Experiment Setup}
\label{sec:exp_setup}
\subsection{Experiment Data}
We pretrain our CSP frontend using unlabeled mixtures and train the entire model in an end-to-end manner using labeled synthetic datasets. Due to the absence of target reference speech in real-world mixtures, it is impossible to assess quality using signal-level metrics. Therefore, following the approach of~\cite{wang2024speech}, we divide experiments in this paper into two parts: synthetic scenario separation experiments and real-world separation experiments. A summary of dataset usage is as follows: we pretrain the frontend with both the LM2Mix~\cite{cosentino2020librimix} and Vox2Mix mixtures, train the separation model using the LM-train-100 set, and test on the test sets of three datasets (LM2Mix, Vox2Mix, and LRS2Mix~\cite{li2022efficient}) for synthetic experiments. For real-world scenario separation experiments, we continue to pretrain with the REAL-M~\cite{subakan2022real} train set and evaluate the separation quality using its test set.  

\subsubsection{Synthetic Scenario Separation Experiments}

We investigate the predictive capability of our CSP frontend by transferring the separation knowledge from the LM2Mix dataset to synthetic Vox2Mix and LRS2Mix datasets.

\textbf{LM2Mix}\footnote{\url{https://github.com/JorisCos/LibriMix}}. The dataset is simulated at a 16 kHz sampling rate based on the LibriSpeech (LS) corpus~\cite{panayotov2015librispeech}. This simulated dataset is divided into four subsets: the LM-train-360 set (50,800 utterances, 921 speakers), the LM-train-100 set (13,900 utterances from 251 speakers), the development set (3,000 utterances from 40 speakers), and the test set (3,000 utterances from 40 speakers). All utterances are randomly selected from the LibriSpeech corpus and mixed at specific loudness levels relative to full scale. The total hours of these subsets are 212h, 58h, 11h, and 11h respectively. 

\textbf{Vox2Mix}. We simulate the two-speaker Vox2Mix dataset at a 16 kHz sampling rate based on the VoxCeleb2 corpus~\cite{chung2018voxceleb2}. Unlike the LM2Mix dataset, the Vox2Mix dataset includes a significant amount of in-the-wild data with diverse accents, nationalities, attenuation levels, noise, and reverberation interference. This diversity makes the Vox2Mix dataset more challenging and representative of real-world scenarios. The simulated database is partitioned into two sets: the Vox-train-53 set (20,000 utterances), and the test set (3,000 utterances). The Vox-train-53 subset consists of utterances that are randomly selected and mixed from the VoxCeleb2 `dev' subset, using specific loudness units similar to those in the LM2Mix. The utterances in the test set are generated from the `test' subset, which is not included in the training data. The total hours of train-53 and test are 53h, and 10h, respectively. 

\textbf{LRS2Mix}. To evaluate the generalization ability of our model comprehensively, we use the test set from the LRS2Mix dataset as the unseen set to evaluate the separation quality. This test set contains 3,000 utterances, which are simulated from the in-the-wild LRS2 dataset~\cite{afouras2018deep}, the spoken sentences recorded by BBC television. The two-speaker mixtures are generated from different speaker audio sources with signal-to-noise ratios ranging from -5 dB to 5 dB. All samples are recorded at a 16 kHz sampling rate and are not involved in either the pretraining or training stage of our model. 

\subsubsection{Real-world Separation Experiments}

In real-world scenario separation experiments, we apply our model to two-speaker mixtures from the REAL-M dataset~\cite{subakan2022real}. This enables us to examine whether our proposed CSP frontend is also suitable to improve the separation quality of the causal separation model in real-world conditions. 

\textbf{REAL-M}. The REAL-M dataset comprises samples collected by asking participants to read a predefined set of sentences simultaneously in various acoustic environments using different recording devices, such as laptops and smartphones. To facilitate causal self-supervised pretraining of our CSP frontend, we partition the 1,436 utterances in the REAL-M dataset into a pretraining set and a test set. The pretraining set consists of 600 mixture samples from the `early collection' subset, while the test set includes 837 samples that are not involved in the pretraining set. It is important to note that all samples in the REAL-M dataset are real recordings of microphones without target reference speech.

\subsection{Model Configuration}
Our CSP frontend is implemented using fairseq\footnote{\url{https://github.com/facebookresearch/fairseq}} toolkit. The causal CNN encoder contains 7 stacked causal CNN blocks. Each block has 512
channels with strides (5,2,2,2,2,2,2) and kernel widths (10,3,3,3,3,2,2). The causal positional embedding layer
has kernel sizes of 128 and is divided into 16 groups. For quantization modules, we utilize $G_u=G_v=2$ codebooks with $R_u=R_v=320$ entries. The Gumbel softmax temperature is annealed from 2 to a minimum of 0.5, with a decay factor of 0.999995 applied at each update. The transformer decoder network comprises 12 stacked transformer decoder layers, with a model dimension of 768, an inner dimension of 3,072, and 8 attention heads. The similarity function $\psi$ is defined as the cosine similarity. We set $K=100$ as the number of clusters and $N=100$ as the number of negative samples.

For the downstream separation tasks, we employ three types of time-domain separation models~\footnote{Note that we present the comprehensive results on ConvTasNet and several key experiments on the others to show the generalizability across different separation models}, ConvTasNet~\cite{luo2019conv}, SkiM~\cite{li2022skim}, and Resepformer~\cite{veluri2023real}, as well as a frequency-domain LSTM model. The implementation of ConvTasNet is based on the Asteroid toolkit\footnote{\url{https://github.com/asteroid-team/asteroid}}, while SkiM and Resepformer are integrated from the official SpeechBrain\footnote{\url{https://github.com/speechbrain/speechbrain}} toolkit implementation into the Asteroid toolkit to ensure a fair comparison across models. All time-domain separation models use a 1-D convolutional layer as the encoder and a 1-D convolutional transpose layer as the decoder, with strides and kernel sizes of 16 and 32, respectively. Other parameters follow the official configurations provided in the Asteroid and SpeechBrain toolkits. For the frequency-domain model, we employ a three-layer LSTM network adapted from the SUPERB benchmark\footnote{\url{https://github.com/s3prl/s3prl}} to estimate the phase-sensitive mask (PSM). The implementation and the checkpoint are available on github\footnote{\url{https://github.com/Wufan0Willan/CSP}}.

\subsection{Training Strategy}
The overall training process of our models is conducted in two distinct stages: the pretraining stage and the training stage. 

\subsubsection{Pretraining stage}
The pretraining strategy for our CSP frontend is outlined as follows: We randomly select 65\% of the frames from all time steps as starting indices and mask the subsequent 10-time steps. Each batch is cropped to a duration of 15.6 seconds. We apply dropout with a rate of 0.1 in the transformer, at the output of the feature encoder, and the input to the quantization module. Additionally, we implement a layer drop for the transformer layers with a rate of 0.05. The learning rate during pretraining is set to 0.0005, and the pretraining terminates when the best validation loss fails to improve for 50 consecutive epochs. We use the Adam optimizer with a weight decay of 0.01 and 32,000 warm-up steps for optimization during pretraining. The temperature parameter in the contrastive loss is set to 0.1.

\subsubsection{Training stage}
To evaluate the compatibility of our CSP frontend, we integrate predictive patterns with both time-domain and frequency-domain separation models. All time-domain separation models are trained using the recipes provided in the Asteroid and Speechbrain toolkits. The batch size for all separation models is 2, with a maximum of 200 training epochs. The learning rate is initialized at 1e-3 for the ConvTasNet and 1.5e-4 for the SkiM and the ReSepformer and is halved if the validation loss does not improve over 5 consecutive epochs. Early stopping is employed if no better model is identified on the validation set for 30 consecutive epochs. The Adam optimizer is utilized for the backpropagation of the separation model only, with the CSP frontend kept frozen during the training stage. For the frequency-domain separation model, we follow the standard procedures outlined in the SUPERB benchmark for speech separation tasks. 

\subsection{Integration with Causal Models}
\begin{figure}[t]
    \centering 
    \includegraphics[width=1.0\linewidth]{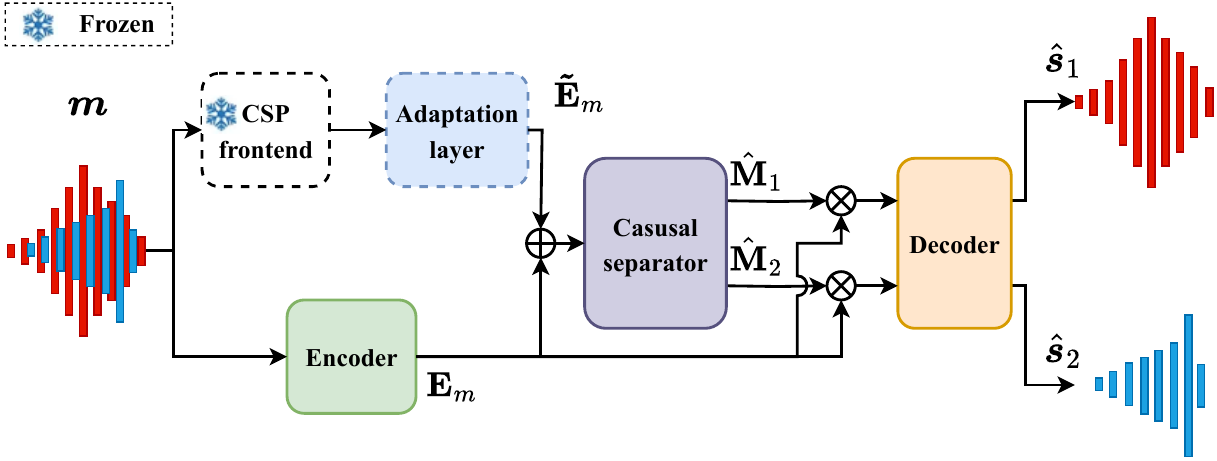}
    \caption{Speech separation training pipeline that incorporates CSP frontends with diverse speech separation models. Here $\boldsymbol{m}$ is the two-speaker synthetic mixture. $\mathbf{E}_m$ is auditory features of mixture waveform. $\mathbf{\tilde{E}}_m$ is the predictive features. $\hat{\mathbf{M}}_1$ and $\hat{\mathbf{M}}_2$ are predicted masks for two speakers, respectively, and $\hat{\boldsymbol{s}}_1$, $\hat{\boldsymbol{s}}_2$ are two reconstructed target reference speech.} 
    \label{Fig.pipeline} 
\end{figure}

We follow the pipeline of~\cite{wang2024speech}, as depicted in Fig.~\ref{Fig.pipeline}, to integrate CSP frontends into different causal separation models. The frontend is pretrained and frozen in the training pipeline. Given a two-speaker mixture waveform $\boldsymbol{m}$, we feed it into both the encoder and the causal frontend with an adaptation layer to extract the auditory embeddings $\mathbf{E}_m$ and the predictive features $\mathbf{\tilde{E}}_m$, respectively. The adaptation layer upsamples the predictive pattern generated from the causal frontend to time-align with the encoder output. Then we sum them together as input to serve as input to the causal separator, which predicts $\hat{\mathbf{M}}_1$ and $\hat{\mathbf{M}}_2$ for each speaker. Finally, the predicted masks are applied to the latent embeddings $\mathbf{E}_m$, and the decoder reconstructs the target speech signal $\hat{\boldsymbol{s}}_1$, $\hat{\boldsymbol{s}}_2$.

\subsection{Evaluation Metrics}

We evaluate the clarity and intelligibility of separated speech with multiple metrics. In the synthetic scenario separation experiments, we use the scale-invariant signal-to-noise ratio improvement (SI-SDRi), signal-to-noise ratio improvement (SDRi), perceptual evaluation of speech quality (PESQ), and Short-Term Objective Intelligibility (STOI) to assess the quality of the separated speech. For the real-world separation experiments, where reference speech for quality evaluation is not available, we utilize the word error rate (WER) of pretrained automatic speech recognition (ASR) systems, and the $\widehat{\text{SI-SNRi}}$ score of the neural SI-SNRi estimator\footnote{\url{https://huggingface.co/speechbrain/REAL-M-sisnr-estimator}}~\cite{subakan2022real} as the measurement metrics.

\section{Results and Analysis}
\label{sec:exp_result}
\subsection{Effectiveness of CSP Frontend}

\begin{table*}
    \centering
    \caption{A comparison among the causal ConvTasNet with our CSP frontend, and its variants.  The $\alpha$ and $\beta$ are scaling factors in the AHP pretext task described in Eq.~\eqref{equ.ahp_hybrid}. The $\gamma$ is the balance factor in Eq.~\eqref{equ.CSP} between AHP and CKD pretext tasks. The development set (Dev) and the test set (Test) are from LM2Mix. `N/A' means not applicable. The `\cmark' indicates a causal model, while a `\xmark' denotes a non-causal model. The \textbf{bold values} are the optimal choice and the `$\uparrow$' means the higher, the better.}
    \addtolength{\tabcolsep}{3.5pt}
    \begin{tabular}{c|c|c|c|c|c|c|c|c|c} 
       \toprule
        \multirow{1}*{System} & Pretrain & Frontend & Separator & \multirow{2}*{$\alpha$} & \multirow{2}*{$\beta$} & \multirow{2}*{$\gamma$} & Dev (LM2Mix) & \multicolumn{2}{c}{Test (LM2Mix)}\\
		(Sys.) & strategy & causality & causality & ~ & ~ & ~ & SI-SDRi (dB) $\uparrow$ & SI-SDRi (dB) $\uparrow$ & SDRi (dB) $\uparrow$ \\

        \midrule 
        1 & N/A &N/A & \xmark & N/A & N/A & N/A & 14.54 & 14.06 & 14.46 \\

        \midrule
        2 & N/A & N/A & \cmark & N/A & N/A & N/A & 9.56 & 8.96 & 9.38 \\

        \midrule
        3 & TD & \multirow{2}*{\cmark} & \multirow{2}*{\cmark} & 1 & 0 & \multirow{2}*{N/A} & 10.91 & 10.49 & 10.88 \\     
        4 & BU & ~ & ~ & 0 & 10 & ~ & 10.47 & 9.97 & 10.37 \\

        \midrule
        5 & \multirow{5}*{AHP} & \multirow{5}*{\cmark} & \multirow{5}*{\cmark} & \multirow{5}*{1} & 100 & \multirow{5}*{N/A} & 10.94 & 10.58 & 10.96\\
   
        6 & ~ & ~ & ~ & ~ & 10 & ~ & 11.27 & 10.86 & 11.25\\
        7 & ~ & ~ & ~ & ~ & 1 & ~ & 11.24 & 10.83 & 11.22 \\
        8 & ~ & ~ & ~ & ~ & 0.1 & ~ & 11.08 & 10.68 & 11.08  \\
        9 & ~ & ~ & ~ & ~ & 0.01 & ~ & 11.16 & 10.77 & 11.16 \\
        
        \midrule 
        10 & CKD & \cmark & \cmark & N/A & N/A & 1 & 10.85 & 10.42 & 10.81\\
        
        \midrule 
        11 & \multirow{5}*{\textbf{Hybrid}} & \multirow{5}*{\textbf{\cmark}} & \multirow{5}*{\textbf{\cmark}} & \multirow{5}*{\textbf{1}} & \multirow{5}*{\textbf{10}} & 100 & 11.46 & 11.04 & 11.43\\
        \textbf{12} & ~ & ~ & ~ & ~ & ~ & \textbf{10} & \textbf{11.62} & \textbf{11.10} & \textbf{11.50}\\
        13 & ~ & ~ & ~ & ~ & ~ & 1 & 11.41 & 10.98 & 11.37 \\
        14 & ~ & ~ & ~ & ~ & ~ & 0.1 & 11.41 & 11.01 & 11.39 \\
        15 & ~ & ~ & ~ & ~ & ~ & 0.01 & 11.38 & 10.97 & 11.36 \\
        
        \midrule
        16 & \multirow{2}*{Hybrid} & \cmark & \multirow{2}*{\xmark} & \multirow{2}*{1} & \multirow{2}*{10} & \multirow{2}*{10} & 15.41 & 15.13 & 15.50 \\
        17 & ~ & \xmark & ~ & ~ & ~ & ~ & 15.75 & 15.52 & 15.92
 \\
       \bottomrule
    \end{tabular}
    \addtolength{\tabcolsep}{-3.5pt}
    \label{tab:baseline}
\end{table*}

We first assess the effectiveness of our CSP frontend on synthetic datasets using conventional signal distortion metrics. The results are presented in Table~\ref{tab:baseline}, where all experiments utilize the causal ConvTasNet~\cite{luo2019conv} as the separation model. We follow the configuration described in~\cite{wang2024speech}, using the mixtures from the LM2Mix train-100 dataset and the Vox2Mix-53h dataset for frontend pretraining. We hypothesize that the CSP frontend is capable of narrowing the performance gap between the causal ConvTasNet and its non-causal counterpart.

The Sys.1 in Table~\ref{tab:baseline} is an offline ConvTasNet model without a pretrained frontend, which achieves an SI-SDRi of 14.06 dB and an SDRi of 14.46 dB on the LM2Mix test set. When the same training methodology is applied to a causal ConvTasNet model (Sys.2) for causal speech separation, there is a notable decline in performance. The separation quality of the causal ConvTasNet drops to 8.96 dB SI-SDRi and 9.38 dB SDRi due to the significant impact of the absence of future information on causal separation systems, as mentioned in Section~\ref{sec:introduction}.

We then investigate the effectiveness of the AHP pretext task introduced in Section.~\ref{sec:AHP}. We individually pretrain our CSP frontend using TD and BU methods. As shown in Sys.3 and Sys.4, the causal separation system with the CSP frontend that pretrained through the TD approach gains 10.49 dB SI-SDRi and 10.88 dB SDRi improvement, while for the BU method, the separated quality is 9.97 dB in SI-SDRi and 10.37 dB in SDRi. Both TD and BU settings outperform the baseline, demonstrating their effectiveness in enabling the frontend model to learn predictive patterns, thereby boosting the separation quality of causal separation models. From Sys.5 to Sys.9, we integrate TD and BU with hyperparameters $\alpha$ and $\beta$ as the AHP pretext task and find that combining these prediction strategies with
appropriate weighting further improves the separation quality. The optimal choice is Sys.6, which brings about 10.86 dB SI-SDRi and 11.25 dB SDRi. 

Next, we examine the impact of the CKD pretext task described in Section~\ref{sec:CKD}. In Sys.10, we utilize the CKD pretext task to pretrain the CSP frontend. It helps the causal ConvTasNet to achieve 10.42 dB SI-SDRi and 10.81 dB SDRi. When both AHP and CKD pretext tasks are combined using the hyper-parameter $\gamma$, as shown from Sys.11 to Sys.15, the separation performance further improves. The optimal value is $\gamma = 10$ in Sys.12, where SI-SDRi and SDRi reach 11.10 dB and 11.50 dB, respectively. These findings demonstrate that the two proposed pretext tasks are beneficial from each other.

Finally, we explore the application of our proposed method within an offline speech separation system. In Sys.16, the integration of our CSP frontend with the offline ConvTasNet results in a separation performance of 15.13 dB in SI-SDRi and 15.50 dB in SDRi. These findings demonstrate that the predictive representations learned by the CSP frontend can effectively enhance separation quality in offline systems. Furthermore, our method is particularly well-suited for pretraining non-causal frontends. As shown in Sys.17, the non-causal frontend pretrained with our proposed pretext tasks achieves an improvement of 1.5 dB in SI-SDRi compared to the baseline offline system (Sys.1) without a frontend.

\subsection{Comparision with Other Pretrained Frontends}
\newcolumntype{Y}{>{\centering\arraybackslash}X}
\newcolumntype{Z}[1]{>{\centering\arraybackslash}p{#1}}
\begin{table*}
    \centering
    \caption{The separation results among the ConvTasNet with various causal pretrained frontends. We test separation quality on the LM2Mix test set and transfer the separation knowledge from the LM2Mix dataset to the Vox2Mix and the LRS2Mix test set without adaptation. The Mix 100h pretrain corpus is the mixture waveform of LM2Mix and Vox2Mix we use in system 11. `N/A' in the following tables means not applicable. `$^\dag$' means we pretrain the frontend with the original receipt using our mixture dataset. The \textbf{bold values} are the optimal choice for the causal speech separation system. The `$\uparrow$' means the higher, the better.}
    \addtolength{\tabcolsep}{-3.5pt}
    \resizebox{0.95\linewidth}{!}{
    \begin{tabularx}{\textwidth}{@{}Z{1.7cm}|Z{2.5cm}|Z{1.5cm}|Y|Y|Y|Y|Y|Y@{}} 
       \toprule
        Separation & Causal & Pretrain & \multicolumn{2}{c|}{Test (LM2Mix)} & \multicolumn{2}{c|}{Test (LM2Mix $\rightarrow$ Vox2Mix)} & \multicolumn{2}{c}{Test (LM2Mix $\rightarrow$ LRS2Mix)}\\
		model & Frontend & Corpus & SI-SDRi (dB) $\uparrow$ & SDRi (dB) $\uparrow$ & SI-SDRi (dB) $\uparrow$ & SDRi (dB) $\uparrow$ & SI-SDRi (dB) $\uparrow$ & SDRi (dB) $\uparrow$\\
        
        \midrule
        \multirow{5}*{ConvTasNet} & N/A & N/A & 8.96 & 9.38 & 4.54 & 4.90 & 2.17 & 2.76 \\
         
        ~ & APC~\cite{chung2020generative} & LS 360h & 8.42 & 8.84 & 4.87 & 5.26 & 4.33 & 4.87\\
        ~ & VQ-APC~\cite{chung2020vector} & LS 360h & 7.73 & 8.15 & 4.39 & 4.78 & 3.81 & 4.36\\
        ~ & Wav2Vec~\cite{schneider2019wav2vec} & LS 960h & 8.77 & 9.18 & 4.47 & 5.01 & 5.28 & 5.65\\
        ~ & VQ-Wav2Vec~\cite{baevski2019vq} & LS 960h & 8.20 & 8.61 & 4.40 & 4.92 & 4.96 & 5.34\\

        \midrule
        \multirow{2}*{ConvTasNet} & $\text{Wav2Vec}^{\dag}$ & Mix 110h & 9.44 & 9.84 & 6.21 & 6.57 & 6.03 & 6.51\\
        ~ & \textbf{CSP} & \textbf{Mix 110h} & \textbf{11.10} & \textbf{11.50} & \textbf{8.08} & \textbf{8.47} & \textbf{6.75} & \textbf{7.18}\\
        
       \bottomrule
    \end{tabularx}
    }
    \addtolength{\tabcolsep}{3.5pt}
    \label{tab:frontend}
\end{table*}
We then assess the domain transfer capability of various causal pretrained frontends, as shown in Table.~\ref{tab:frontend}, using causal ConvTasNet as the separator. Compared to the baseline model without any frontend, integrating conventional causal frontends slightly improves separation quality on out-of-domain Vox2Mix and LRS2Mix test sets, but results in decreased performance on the in-domain LM2Mix test set. This decline may be due to the input mismatch between the pretraining corpus and the finetuning dataset, as these frontends are pretrained on single-speaker reference speech but finetuned on multi-speaker mixtures. To verify this hypothesis, we further utilize the Mix 110h dataset\footnote{The `Mix 110h' pretrain corpus contains 53h mixtures from the Vox-train-53 subset and 58h mixtures from the LM-train-100 subset.} to pretrain the Wav2Vec with the original receipt, denoted as Wav2Vec$^\dag$, to make a fair comparison with our CSP frontend. The ConvTasNet with the Wav2Vec$^\dag$ frontend shows improvement on both in-domain and out-of-domain test sets when compared with the baseline model. In contrast, our proposed CSP frontend significantly improves separation quality for the causal ConvTasNet. Specifically, it brings about 2.14 dB SI-SDRi and 2.12 dB SDRi on the in-domain LM2Mix test set. For out-of-domain scenarios present during pretraining, such as Vox2Mix, the CSP frontend delivers improvements of 3.54 dB SI-SDRi and 3.57 dB SDRi. Notably, even for the LRS2Mix dataset, which is unseen during pretraining or fine-tuning, the CSP frontend achieves improvements of 4.58 dB SI-SDRi and 4.42 dB SDRi. These results highlight the effectiveness of our CSP frontend in transferring separation knowledge from in-domain datasets to out-of-domain test sets.

\begin{figure}[t]
	\centering 
        \includegraphics[width=0.95\linewidth]{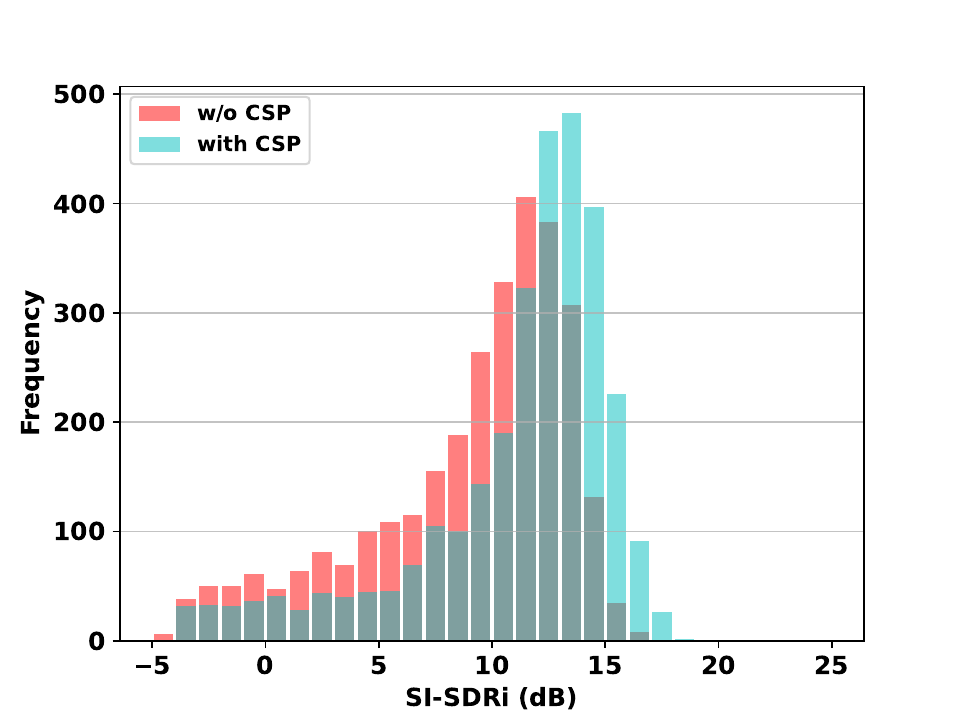}
	\caption{The histogram of SI-SDRi (dB) on individual utterances. The red bar is the number of samples for special SI-SDRi (dB) using the causal ConvTasNet model. The blue bar on the causal ConvTasNet with our proposed CSP frontend in Sys.11.} 
	\label{Fig.LM2Mix} 
\end{figure}

In Fig.~\ref{Fig.LM2Mix}, we analyze the improvement in SI-SDRi for individual utterances. The red bars represent samples from the causal ConvTasNet baseline without any frontend, where the majority achieve SI-SDRi above 5 dB, though a significant number remain below 0 dB. In contrast, the blue bars represent samples processed with our CSP frontend, leading to substantial improvements in SI-SDRi scores, with most surpassing 10 dB. These findings indicate that integrating our CSP frontend enables the causal ConvTasNet to produce higher-quality samples.

\subsection{Results using Different Separation Models}

\begin{figure}[t]
	\centering 
\includegraphics[width=0.75\linewidth]{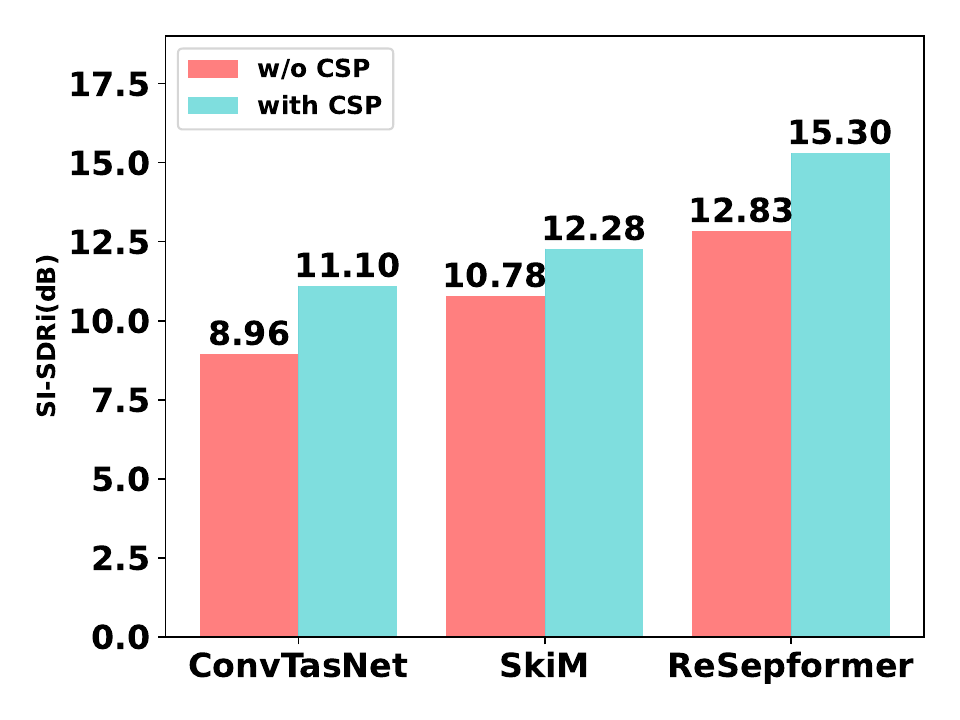}
	\caption{The SI-SDRi (dB) of different separation models to separate mixtures from the LM2Mix test set. The red bar is the separation model without any pretrained frontend and the blue bar is the counterpart with our CSP frontend.} 
	\label{Fig.separator} 
\end{figure}

The experiments in previous sections convincingly demonstrate the effectiveness of our CSP frontend on the causal ConvTasNet. To measure the adaptability of the causal pretrained frontend, we employ both state-of-the-art time-domain and frequency-domain separation models to evaluate the separated speech quality on the LM2Mix test set. 

We start from three time-domain separation models: ConvTasNet~\cite{luo2019conv}, SkiM~\cite{li2022skim}, and ReSepformer~\cite{libera2024resource}. As depicted in  Fig.~\ref{Fig.separator}, all separation models integrated with our proposed CSP frontend (blue bars) demonstrate superior performance compared to their counterparts without a frontend (red bars). Compared to baseline models without a frontend, integrating our proposed CSP frontend with SkiM achieves an additional 1.50 dB improvement in SI-SDRi. For the state-of-the-art ReSepformer, the CSP frontend boosts SI-SDRi by 2.47 dB.

\begin{table}
	\caption{Study of Real-time factor (RTF) and latency evaluation for causal ConvTasNet model on the LM2Mix test set with an Intel(R) Xeon(R) E5-2640 v4 2.40GHz CPU. `N/A' means not applicable.}
	\label{tab:cpu}
	\centering 
        \resizebox{\linewidth}{!}{
	\begin{tabular}{c|c|c|c|c|c}
		\toprule
		Separation & \multirow{2}*{Frontend} & Ideal latency & MACs & \multirow{2}*{RTF} & Latency\\
		model & ~ & (ms) &(G/s) & ~ & (ms)\\

            \midrule
		\multirow{2}*{ConvTasNet} & N/A & 1 & 4.6 & 0.63 & 1.63\\ 
        ~ & CSP & 20 & 26.6 & 0.81 & 36.20 \\
                \midrule
		\multirow{2}*{ReSepformer} & N/A & 150 & 5.8 & 0.32 & 198.00 \\ 
        ~ & CSP & 150 & 16.6 & 0.65 & 247.50 \\
		\bottomrule
	\end{tabular}
        }
\end{table}

To evaluate the feasibility of deploying our CSP frontend partial causal separation model (ReSepformer) for real-time processing, we measure the multiplier–accumulator operations (MACs) per second\footnote{\url{https://github.com/zhijian-liu/torchprofile}}, the real-time factor (RTF), and the latency of both causal (ConvTasNet) and partial causal (ReSepformer) separation model on the LM2Mix test set. These tests are conducted using an Intel(R) Xeon(R) E5-2640 v4 2.40GHz CPU, restricted to a single thread. As shown in Table.\ref{tab:cpu}, incorporating the CSP frontend increases MACs. This suggests that the frontend is suitable for high-performance cloud systems rather than resource-constrained edge devices. Additionally, the frontend contributes to higher latency due to its large stride size in causal separation models. For partially causal models, however, latency primarily comes from the separator itself. Nevertheless, the RTFs of all models remain below 1.0, indicating that the models with the frontend are capable of streaming decoding. 

\subsection{Evaluation on SUPERB Benchmark}
We conduct separation experiments using the official SUPERB benchmark to evaluate the effectiveness of predictive patterns with a compact parameter configuration. To achieve this goal, we replace the original non-causal three-layer Bi-LSTM separation model with a causal three-layer LSTM model to predict the PSM for the target reference speech in a casual setting. As shown in Table.~\ref{tab_freq}, the results align with the trends observed in the time-domain experiments: pretraining on mixtures yields a modest improvement in separation quality, while our proposed CSP frontend achieves the highest scores in PESQ, STOI, and SI-SDRi metrics. Notably, the improvement in SI-SDRi is more pronounced than in PESQ and STOI, suggesting that the CSP frontend offers significant benefits for signal quality over speech coherence.

\begin{table}
	\caption{Experiment results on the LM2Mix dataset with frequency-domain separation model. The LSTM-based separation model predicts the PSM of the mixture to reconstruct the target reference speech. `$^\dag$' means we pretrain the frontend with the original receipt using our mixture dataset. The \textbf{bold values} are the optimal choice for the causal speech separation system. The `$\uparrow$' means the higher, the better.}
	\label{tab_freq}
	\centering 
	\begin{tabular}{c|c|c|c|c}
		\toprule
		Separation & \multirow{2}*{Frontend} &  \multicolumn{3}{c}{Test (LM2Mix)} \\
		model & ~ & PESQ $\uparrow$ & STOI $\uparrow$ & SI-SDRi (dB) $\uparrow$\\
		\hline
  
		\multirow{6}*{LSTM} & STFT & 1.40 & 0.80 & 6.30\\ 
        ~ & APC & 1.43 & 0.81 & 6.78\\
        ~ & VQ-APC & 1.40 & 0.80 & 6.24\\
	~ & Wav2Vec & 1.42 & 0.83 & 7.07\\
        ~ & VQ-Wav2Vec & 1.35 & 0.80 & 6.07\\ 
        ~ & Wav2Vec\dag & 1.53 & 0.85 & 8.68\\
        ~ & \textbf{CSP} & \textbf{1.56} & \textbf{0.88} & \textbf{9.02}\\
		\bottomrule
	\end{tabular}
\end{table}

\subsection{Evaluation on Real Dataset}
Previous experiments consistently demonstrate that our proposed CSP frontend effectively addresses the causal separation, facilitating the transfer of separation knowledge from labeled datasets to unlabeled test sets in causal separation models. In this section, we evaluate our model on real two-speaker mixtures from the REAL-M dataset (excluding samples from the `early collection' set) to confirm the advantages of mitigating the domain gap for achieving high-quality speech separation in real-world scenarios. We utilize two pretrained ASR models—an E2E transformer and a WavLM-based transformer from the ESPNet toolkit~\cite{watanabe2018espnet}—to assess the quality of the separated speech indirectly.

Table.~\ref{tab_asr} explores two configurations of our proposed CSP frontend: `CSP' represents the pretrained frontend in system 11, while `CSP+' denotes the frontend that continually pretrained with 600 mixtures from the `early collection' set of REAL-M dataset. Notably, leveraging our CSP frontend significantly improves the separated speech quality, consequently reducing the WER of the ASR system. For example, the ReSepformer without our CSP frontend achieves 57.75\% WER on the REAL-M dataset. With the `CSP' frontend, the WER of the ReSepformer is reduced by 14.85\% WER (42.90\%). The incorporation of additional real mixtures for frontend pretraining further leads to a 2.72\% WER improvement (40.18\%). Similar trends are observed for the ConvTasNet and the SkiM separation models. These results demonstrate that alleviating the domain gap between real and synthetic samples through our proposed CSP frontend is effective for real-scenario speech separation models to generate high-quality separated speech.

\begin{table}
    \caption{A comparison among different separation models with our CSP frontend to transfer the separation knowledge from the LM2Mix dataset to the REAL-M dataset. We implement two types of ASR models to evaluate with the WER metric. We also employ the SI-SNRi neural estimator~\cite{subakan2022real} to evaluate the separation quality. The \textbf{bold values} are the optimal choice for the causal speech separation system. The `$\uparrow$' means the higher, the better, while the `$\downarrow$' means the lower, the better.}
    \centering
    \begin{tabular}{c|c|c|c|c} 
       \toprule
        Separation & \multirow{2}*{Frontend} & E2E ASR & WavLM ASR & $\widehat{\text{SI-SNRi}}$\\
        model & ~ & WER (\%) $\downarrow$ & WER (\%)$\downarrow$ & (dB)$\uparrow$\\

        \midrule    
        \multirow{3}*{ConvTasNet} & N/A & 88.64 & 66.91 & 0.45 \\
        ~ & CSP & 71.36 & 48.94 & 0.79 \\
        ~ & \textbf{CSP+} & \textbf{67.89} & \textbf{45.08} & \textbf{0.92} \\

        \midrule    
        \multirow{3}*{SkiM} & N/A & 85.34 & 59.97 & 0.59 \\
        ~ & CSP & 73.40 & 49.04 & 0.81 \\
        ~ & \textbf{CSP+} & \textbf{70.63} &\textbf{45.54} & \textbf{0.84} \\

        \midrule    
        \multirow{3}*{ReSepformer} & N/A & 84.26 & 57.75 & 0.74  \\
        ~ & CSP & 64.69 & 42.90 & 1.21\\
        ~ & \textbf{CSP+} & \textbf{61.54} & \textbf{40.18} & \textbf{1.23}\\
       \bottomrule
    \end{tabular}
    \label{tab_asr}
\end{table}

We also employ the neural SI-SNRi estimator~\cite{subakan2022real} for perceptual quality measurement to evaluate our models.  In contrast to other objective metrics that rely on the availability of the target reference speech, the neural SI-SNRi estimator directly utilizes the $\widehat{\text{SI-SNRi}}$ score to approximate the SI-SNRi value. However, since the estimator is trained on the 8 kHz WHAMR! dataset with reverberation, it cannot accurately estimate the SI-SNRi scores for the 16 kHz test sets. Despite this limitation, we observed consistent relative improvements in all three time-domain causal separation models when integrated with our CSP frontend. In addition, the extra real mixtures in the pretraining corpus slightly improve the perceptual quality of separated speech as shown in the CSP+ setting when compared with the CSP frontend. 

\section{Conclusion}
\label{sec:con_ssl}

In this work, we introduce a causal self-supervised pretrained (CSP) frontend to tackle the challenge of `\emph{causal separation}'.  By developing two innovative pretext tasks—autoregressive hybrid prediction (AHP) and contextual knowledge distillation (CKD), we demonstrate that predictive patterns learned from unlabeled mixture data can effectively compensate for the absence of future information in causal systems. Experiments demonstrate significant improvements across various models and metrics, with optimal performance achieved when these two pretext tasks are properly integrated. Moreover, our approach highlights an underexplored direction for mitigating domain mismatch in streaming applications through self-supervised learning from unlabeled real-world scenario mixtures.

\bibliographystyle{IEEEbib}
\bibliography{IEEEabrv,Bibliography}

\vfill

\end{document}